# Superconductor/Ferromagnet Heterostructures: A Platform for Superconducting Spintronics and Quantum Computation

*Ranran Cai[1,2]\*, Igor Žutić[3]\*, Wei Han[1]\**

[1]International Center for Quantum Materials, School of Physics, Peking University, Beijing 100871, P. R. China

[2]CAS Key Laboratory of Quantum Information, University of Science and Technology of China, Hefei, Anhui 230026, P. R. China

[3]Department of Physics, University at Buffalo, State University of New York, Buffalo, New York 14260, USA

\*Correspondence to: cairanran@ustc.edu.cn; zigor@buffalo.edu; weihan@pku.edu.cn

The interplay between superconductivity and ferromagnetism in the superconductor/ferromagnet (SC/FM) heterostructures generates many interesting physical phenomena, including spin-triplet superconductivity, superconducting order parameter oscillation, and topological superconductivity. The unique physical properties make the SC/FM heterostructures as promising platforms for future superconducting spintronics and quantum computation applications. In this article, we review important research progress of SC/FM heterostructures from superconducting spintronics to quantum computation, and it is organized as follows. Firstly, we discuss the progress of spin current carriers in SC/FM heterostructures including Bogoliubov quasiparticles, superconducting vortex, and spin-triplet Cooper pairs which might be used for long-range spin transport. Then, we will describe the π Josephson junctions and its application for constructing π qubits. Finally, we will briefly review experimental signatures of Majorana states in the SC/FM heterostructures and the theoretically proposed manipulation, which could be useful to realize fault-tolerant topological quantum computing.

**Keywords:** superconductor/ferromagnet heterostructures, superconducting spintronics, spin-triplet superconductivity, π Josephson junction, π Qubit, Majorana fermions.

## 1. Introduction

Superconductivity and ferromagnetism usually compete with each other; thus, they rarely co-exist in one material. For conventional BCS superconductor (SC) [1], the electron-phonon interaction induces the pairing between spin-up and spin-down electrons to form spin-singlet Cooper pairs. On the other hand, for a typical ferromagnet (FM), the spins are aligned along the same direction due to the exchange coupling. For the last several decades, the quest for the coexistence of SC and FM orders is of great fundamental physical significance, which has drawn extensive attention in condensed matter and materials physics. In 1964, Fulde, Ferrell [2] and Larkin, Ovchinnikov [3] (FFLO) have independently considered the superconducting pairing in weakly ferromagnetic materials or under an external magnetic field, which predicted the existence of the superconducting phase oscillations and spin-triplet Cooper pair components. Inspired by the pioneering FFLO work, researchers have tried to introduce the superconducting pairing into FM via proximity effect [4]. In the SC/FM heterostructures, the interplay between superconductivity and ferromagnetism gives rise to many extremely interesting phenomena. For example, the infinite magnetoresistance (MR) was discovered in superconducting spin-valve devices, which has promoted the rapid progress of the superconducting spintronics field [5]. One major research purpose of spintronics is to manipulate the degree of spin within its dephasing time for fabricating the high-speed spin-based logic devices with low-power consumption [6,7]. To meet such a goal, researchers have spent a significant effort searching for the new-type of spin-current carriers in SC/FM heterostructures. For example, SC quasiparticles [8] with charge-spin separation characteristic and topologically-protected vortices [9] were proposed to sustain spin current over long distances and lifetimes. Even more exciting, the spin-triplet Cooper pairs [10,11] can be constructed at the SC/FM interface to potentially mediate the dissipationless spin current. Then, by considering the superconducting phase oscillation effect, π FM

Josephson junctions can be realized in SC/FM heterostructures. The advances of superconducting spintronics also indicates the further potential application of SC/FM heterostructures in quantum computation. For example, the π FM Josephson junctions as a π phase shifter can be used to construct π qubit [12,13] with possible long-coherence times [14] and simplified circuit structure [15,16]. In addition, *p*-wave SC, as orbit counterpart of spin-triplet SC, under topological nontrivial phase [17] are theoretically expected to host the Majorana states [18] and to potentially realize fault-tolerant topological quantum computation [19]. Recently, both the zero-dimensional (0D) Majorana bound states and one-dimensional (1D) chiral states have been explored in several SC/FM heterostructures [20-23]. This remarkable progress shows that the SC/FM heterostructures can be an excellent platform to explore the interplay between the competing orders of superconductivity and ferromagnetism, and an ideal system for the superconducting spintronics and quantum computation applications.

In the paper, we aim to review the key progress along the roadmap from superconducting spintronics to quantum computation, as illustrated in the Fig. 1. Following the roadmap, the review is divided into two parts. First, we discuss the research progress of SC/FM heterostructures in superconducting spintronics including Bogoliubov quasiparticles, superconducting vortex, spin-triplet SC and π FM Josephson junctions. In the second part for quantum computation applications, π Josephson junctions-based SC qubits will be first discussed. Then, the experimental progress and theoretical manipulation of Majorana states in the SC/FM heterostructures will be briefly reviewed. At the end, an outlook of the SC/FM heterostructures is discussed from superconducting spintronics to quantum computation.

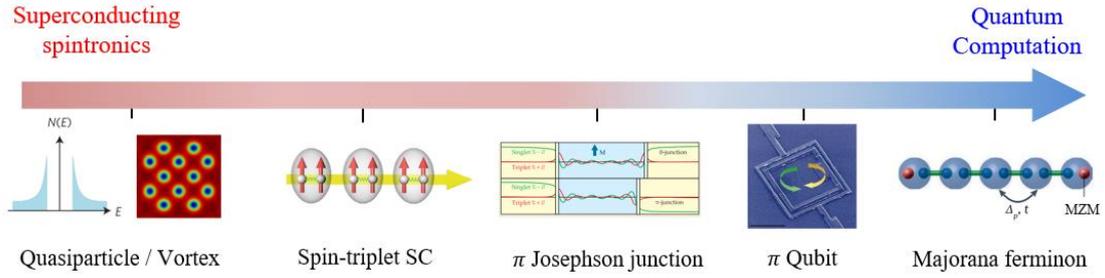

**Figure 1. The research progress of SC/FM heterostructure along the roadmap from superconducting spintronics to quantum computation.** From left to right, the schematics of Bogoliubov quasiparticles, superconducting vortex, spin-triplet SC, π Josephson junctions, SC qubits and Majorana zero modes (MZM). These figures are adapted from Ref [11], with permission, copyright Springer Nature 2015; Ref [24], with permission, copyright Wiley 2020; Ref [25], with permission, copyright Springer Nature 2020; Ref [26], with permission, copyright American Institute of Physics 2010; Ref [27], with permission, copyright Springer Nature 2008; Ref [17], with permission, copyright IOP publishing 2001, receptively.

## 2. SC/FM heterostructure for superconducting spintronics

### 2.1. SC quasiparticles for superconducting spintronics

The early research in superconducting spintronics mainly rely on the spin polarization of the SC quasiparticles, since it is predicted to be an ideal medium with long spin lifetime ($\tau_s$) [8,28]. SC quasiparticles, referred to the Bogoliubov quasiparticles [Fig. 2(a)], are low energy single fermion excitations in SCs, and can be viewed as the coherent superpositions of electrons and holes [29]. Determined by the excitation energy ($E$), the relative components of electron and hole in SC quasiparticles will be changed. Close to the SC energy gap edge ($E \cong \Delta_{sc}$), the electron and hole contents are almost the same, so that the Bogoliubov quasiparticles show charge neutrality. As the $E$ increases to a much larger energy compared to $\Delta_{sc}$, SC particles behave similarly as the ordinary electrons ($E \gg \Delta_{sc}$) and holes ($E \ll -\Delta_{sc}$). Therefore, the density of state (DOS) of SC quasiparticles exhibits a maximum at the SC energy gap edge and decreases as the excitation energy further increases, as shown in Fig. 2(a). However, unlike their charge, SC quasiparticles always retain spin-1/2 property, regardless of their excitation energy. Theoretically, the charge-spin separation of SC quasiparticles [30] can decrease the spin-orbit scattering effect to enhance $\tau_s$ [31]. Another mechanism for the long $\tau_s$ is related to the smaller group velocity of the SC quasiparticles near the SC energy gap edge [11,32] due to the nearly flat band structure.

The experimental observation of extremely long $\tau_s$ of SC quasiparticles has been first demonstrated by Yang et.al. in superconducting Al thin films [8]. As shown in Fig. 2(b), the devices consist of a bottom ferromagnetic CoFe layer whose magnetization is pinned by the antiferromagnetic IrMn via exchange bias, an upper CoFe as free layer, and an insulating MgO layer with thin Al layer buried inside it. A small external magnetic field can change the configuration of the two CoFe layers' magnetization between the parallel and antiparallel states. $\tau_s$ of SC quasiparticle can be deduced from the measured tunneling magnetoresistance ratio at various bias voltages. As shown in Fig. 2(c), $\tau_s$ of SC quasiparticles exceed 0.1 ms below the superconducting critical temperature ($T_C$), which is about $10^6$ lager than those of ordinary electrons in Al. The strong temperature dependence of $\tau_s$ is consistent with the theoretical prediction of SC quasiparticles as a long lifetime spin-current medium [32]. Similar charge-spin separation property enhanced $\tau_s$ in superconducting Al has also been reported by Quay et al. based on lateral spin valve structure [33]. In addition to the long spin lifetimes, a giant inverse spin Hall effect has been demonstrated in superconducting NbN, which could be important for efficient charge-spin conversion applications [28]. These experimental breakthroughs of SC quasiparticles might pave the way for future spin-based memories or logic devices [34].

Recently, the dynamical spin current mediated by SC quasiparticles has been investigated both theoretically and experimentally. A large coherence peak of the Gilbert damping in GdN/NbN/GdN heterostructures was observed via ferromagnetic resonance (FMR) method [35]. The large enhanced Gilbert damping represents an enhanced dynamical spin susceptibility slightly below $T_C$, where DOS of SC quasiparticle accumulation is calculated by Inoue et. al. [36]. Moreover, using the inductive detection of magnetization dynamic [37], Müller et al. have successfully separated the damping-like torque, generated by the quasiparticle inverse spin Hall effect, from field-like torque [38].

## 2.2. Superconducting vortex for superconducting spintronics

Unlike the quasiparticle as an intrinsic single-particle excitation in SC, the superconducting vortex refers to the quantum magnetic flux formed in type-II SC due to the magnetic field penetration, which was first predicted by Abrikosov [39]. However, such a vortex has not been considered in the field of superconducting spintronics until recently, Kim et al. proposed that the superconducting vortex can be a robust spin current medium because its vorticity is topologically protected [9]. The vorticity is determined by the supercurrent flow direction along the superconducting vortex edge, which can be expressed by the following equation:

$$q = \frac{1}{2\pi} \oint d\mathbf{r} \cdot \nabla \phi, \tag{1}$$

where $\phi$ is the phase of the superconducting order parameter winding along the edge circle of the vortex. As shown in Fig. 2(d), the spin angular momentum and vorticity can be transformed into each other at the FM/SC interfaces with interfacial spin Hall effect and/or other spin-orbit coupling (SOC) effects. At the left FM/SC interface, the precession of magnetic momentum in FM layer induces spin-current transport across the interface, which can be transformed into charge current and exerts the transverse Lorentz force on SC to drive the vortex motion. At the right FM/SC interface, vortex in SC can also be transformed into spin angular momentum in FM as a result of the Faraday law and inverse spin Hall effect. Due to the topologically-nontrivial vorticity [9], the superconducting vortex coded with spin information can be used to realize the long-range spin-transport applications.

In addition to mediating spin transport directly, superconducting vortex can also be used to regulate spin-wave propagation in SC/FM heterostructures via the vortex-magnon interaction. In the Shubnikov phase [40], SC vortex forms lattice structure, which can act as a magnonic crystal with spatially periodically modulated magnetic fields. Hence, the magnon spin current flowing in the vortex lattice will be modulated and results in the magnon dispersion spectrum. As shown in Fig. 2(e), Dobrovolskiy et al. fabricated the Py/Nb heterostructures with a 5 nm Pt spacing layer [41]. The magnon spin current is excited by oscillating Oersted field by antenna (port 1), and is detected by the right antenna (port 2) after the spin wave current propagates through the Py wave guide. To modulate the magnon spin current, the applied magnetic field consists of both in-plane ($\mu_0 \boldsymbol{H}_\parallel$) and out-of-plane ($\mu_0 \boldsymbol{H}_\perp$) components. The $\mu_0 \boldsymbol{H}_\parallel$ component is used to set the magnetization direction of Py along its long axis and to define the magnon spin-wave spectrum. The $\mu_0 \boldsymbol{H}_\perp$ component is used to modulate the vortex lattices constant ($a_{vL}$). When the magnon spin-wave vector ($k_{sw}$) and the vortex lattice wavenumber ($k_{vL}$) satisfy Bragg scattering condition:

$$2k_{vL} = nk_{sw}, \tag{2}$$

the transmission spin-wave spectrum will strongly be modulated by the superconducting vortex lattices. Figure 2(f) shows the phase diagram of normalized spin-wave transmission through the Py/Nb heterostructures as a function of $\mu_0 \boldsymbol{H}_\perp$ and spin-wave frequency. At the first ($n$ = 1) and second ($n$ = 2) Bragg scattering conditions, there is a significant spin-wave absorption indicating that the magnon spin wave can be effective modulated by superconducting vortex lattice. Beyond this, the interaction between the moving vortex lattice and magnon spin wave can also induce the absorption frequency shift in transmission spectrum due to Doppler effect [41,42].

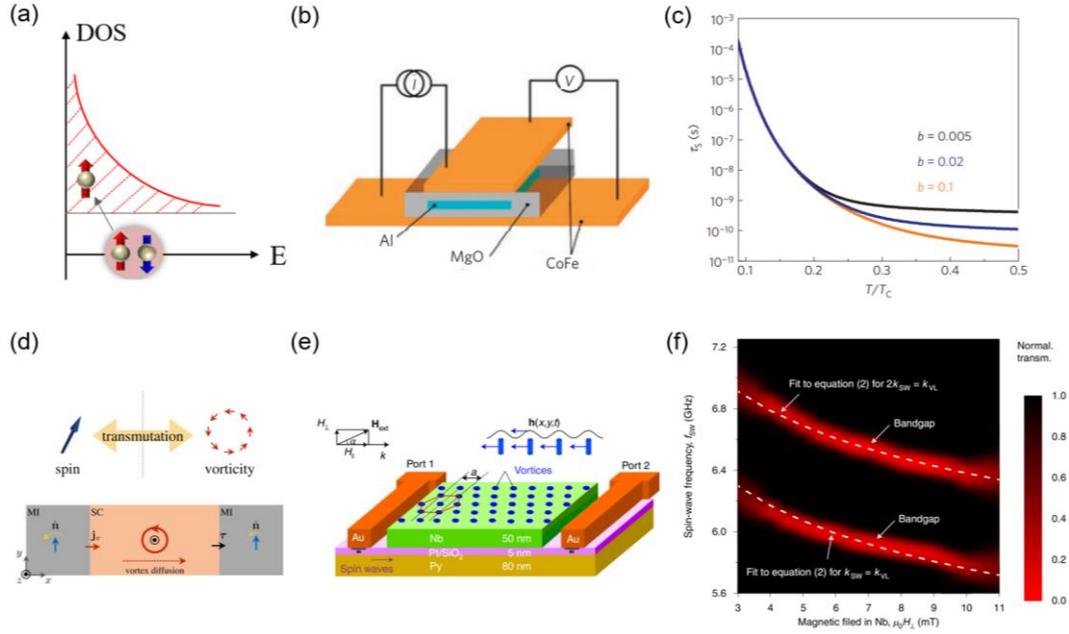

**Figure 2. SC quasiparticles and vortex mediate the spin current.** (a) Schematic of the density of state (DOS) of Bogoliubov quasiparticles in a SC. (b) Schematic of magnetic tunnel junction made of CoFe/MgO (Al)/CoFe and the measurement setup to probe the spin lifetimes in Al. (c) The temperature dependence of the spin lifetime deduced from tunneling magnetoresistance (TMR) ratio with various spin-orbit coupling strength parameter (b = 0.1, 0.02, and 0.005, respectively). (d) The schematic of spin current mediated by SC vortex. Upper plane: the transmutation between the spin angular momentum and vorticity of SC vortex. Lower plane: The theoretical prediction of spin transport via SC vortex liquid in the magnetic insulators (MI)/SC/MI structures. (e) The device structure of Nb/Py heterostructures to probe the coupling between magnon and vortex. The Au electrodes are used as antennas to excite and detect the magnon spin wave in Py. (f) The normalized magnon spin-wave transmission diagram as a function of out-of-plane magnetic field and spin-wave frequency. The two bandgap signatures agree well with the first and second Bragg scattering conditions. (b-c) are adapted from Ref [8], with permission, copyright Springer Nature 2010. (d) is adapted from Ref [9], with permission, copyright APS 2018. (e-f) are adapted from Ref [41], with permission, copyright Springer Nature 2019.

### 2.3. Spin-triplet SC

The search of dissipationless spin current is one of most challenging but exciting directions in the field of spintronics. Cooper pairs in SC dominate dissipationless charge current, which are considered for carrying dissipationless spin current if the Cooper pairs can carry spin angular momentum. However, for conventional BCS SC, the spin-singlet Cooper pairs carry zero-spin angular momentum and cannot transmit spin current since the odd parity of pairing between spin-up and spin-down electrons ($|\uparrow\rangle|\downarrow\rangle - |\downarrow\rangle|\uparrow\rangle$). On the other hand, equal-spin-triplet Cooper pairs ($|\uparrow\rangle|\uparrow\rangle$ or $|\downarrow\rangle|\downarrow\rangle$), pairing between the electrons with same spin, might hold both dissipationless charge and spin current.

Despite the rareness of spin-triplet SC in the nature [43-45], the spin-triplet Cooper pairs can be constructed at the SC/FM interface due to the interplay between superconductivity and ferromagnetism [11]. Detailed discussion of spin-singlet and spin-triplet pairing conversion via spin mixing and spin rotation processes has been reviewed in previous articles (Refs. [4,10]). The equal-spin-triplet SC can be generated as a result of the real space spin texture at the SC/FM interface, including spin active interfaces [46-50], artificial magnetic multilayer [51,52], and noncollinear spin structure in topological magnets [53].

In contrast to spin-singlet SC, equal-spin-triplet SC is not limited by ferromagnetic polarization to induce superconducting proximity effect into FM and form long-range Josephson coupling in FM, as illustrated in Fig. 3(a). Therefore, the experimental probe of the spin-triplet SC has been intensively performed using the Josephson current technique. Keizer et al. were the first to report the long-range Josephson supercurrent in NbTiN/$CrO_2$/NbTiN lateral Josephson devices in 2006 (inset of Fig. 3(b)), where the $CrO_2$ is a half metal with nearly 100% spin polarization and NbTiN is a conventional *s*-wave SC [46]. As shown in Fig. 3(b), Josephson supercurrent can flow across the 310 nm $CrO_2$ spacer between the two NbTiN superconducting electrode at $T$ = 1.6 K. Such a long distance significantly exceeds the coherence length of conventional BCS SC Cooper pair in half metal, which can be most likely explained by the spin-triplet supercurrent. Subsequently, the long Josephson coupling in $CrO_2$ has also been reported by other researchers [54,55]. In these works, equal-spin-triplet Copper pairs are claimed to be converted from spin-singlet Copper pairs due to the spin-active interface of $CrO_2$/NbTiN [56]. Furthermore, by inserting Ni thin film into the interface of $CrO_2$/SC, equal-spin-triplet SC can be controllable via changing the magnetization direction of Ni [57,58]. Recently, Sanchez-Manzano et al. realized the high-temperature spin-triplet Josephson supercurrent at YBCO/half metal (LSMO) heterostructures [48]. As shown in Fig. 3(c), the Josephson supercurrent persist up to 40 K across the 1 μm LSMO spacer. In addition to the half-metallic systems, artificial magnetic multilayer with noncollinear spin texture have also been explored experimentally to support long range spin-triplet supercurrents [51,52,59]. Furthermore, natural spin textures in FM, such as magnetic domain wall [60] or topological Kagome magnetic material $Mn_3Ge$ with noncollinear spin structure [53], can also be used to support spin-triplet Josephson coupling.

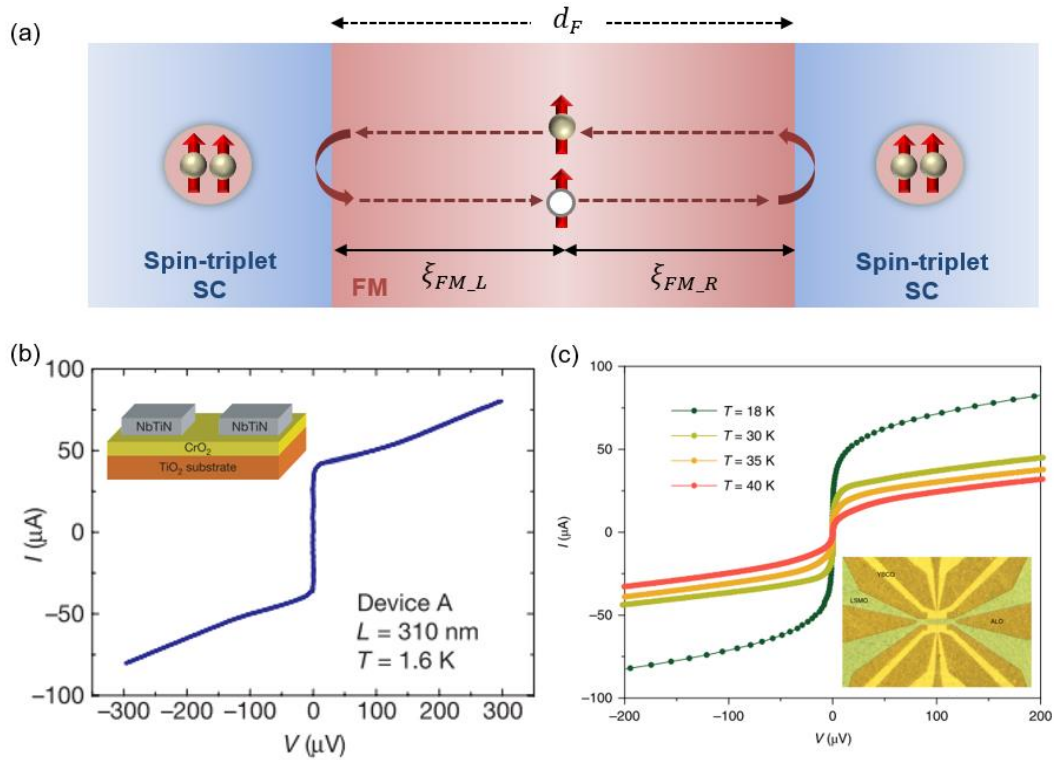

**Figure 3. Experimental progresses of long-range spin-triplet Josephson supercurrent in SC/FM heterostructures.** (a) The schematic of spin-triplet Cooper pairs carries a long-range Josephson supercurrent across the FM layer. (b) The long-range spin-triplet supercurrent in $NbTiN/CrO_2/NbTiN$ Josephson junction at $T = 1.6$ K. Inset: the schematic of $CrO_2$ Josephson device. (c) Josephson effect in high-temperature SC YBCO/half metallic LSMO heterostructure from $T = 18$ K to $T = 40$ K. Inset: the optical image of the typical device with spacing of 1 μm between two high-$T_C$ SC electrodes. (b) are adapted from Ref [46], with permission, copyright Springer Nature 2006. (c) is adapted from Ref [48], with permission, copyright Springer Nature 2022.

Apart from real-space spin texture, SOC or other spin-momentum locking effect can intrinsic induce the spin texture in the *K* space. Several decades ago, Gor'kov and Rashba have pointed out that superconducting paring in such spin-lifting systems is a mixing of both spin-singlet and spin-triplet components [61]. To realize and detect the SOC generated equal-spin-triplet Cooper pairs, Cai et al. fabricated the SC/FM heterostructure devices based on the van der Waals (vdWs) ferromagnet Fe intercalated $TaS_2$ ($Fe_{0.29}TaS_2$) thin flake and measured the compelling evidences of spin-triplet Andreev reflection at the Rashba interface [62]. Unlike the conventional Andreev reflection [63], equal-spin-triplet Andreev reflection referring the incident electron and reflected hole with the same spin-polarized direction [49]. This experiment was motivated by a theoretical work proposed by Hogel et al. [64]. At the interface, the spin-rotation symmetry broken leads to the spin-singlet paring with a spin-triplet component, which acts as spin-mixing processes. Then, the FM magnetization sets the spin-quantization axis, and non-spin-polarized spin-triplet component ($|\uparrow\rangle|\downarrow\rangle + |\downarrow\rangle|\uparrow\rangle$) can be projected onto the spin-quantization axis, which is considered as the spin-rotation process. For example, for FM magnetization along *z* axis (perpendicular to the interface), non-spin polarized spin-triplet Cooper pairs ($|\uparrow\rangle_x|\downarrow\rangle_x + |\downarrow\rangle_x|\uparrow\rangle_x$ and $|\uparrow\rangle_y|\downarrow\rangle_y + |\downarrow\rangle_y|\uparrow\rangle_y$ can be projected to the spin-quantization axis as equal-spin-triplet Cooper pair $|\uparrow\rangle_z|\uparrow\rangle_z$ and $|\downarrow\rangle_z|\downarrow\rangle_z$ [62]. Due to the in-plane spin-momentum locking of Rashba SOC, the spin-triplet Andreev reflection possibility will be highly anisotropic that is dependent on the spin-polarization direction of electrons in the ferromagnetic $Fe_{0.29}TaS_2$. Experimentally, the interfacial resistance ($R_{3T}$) is proportional to the spin-triplet Andreev reflection efficiency, and is measured via three-terminal measurement. At *T* = 2 K, by changing the spin-polarization direction of the electrons in $Fe_{0.29}TaS_2$ via external magnetic field [Fig. 4(b)], the interfacial resistance shows two-fold symmetry, which agrees well with theoretical perdition of spin-triplet MR, as shown in Fig. 4(b) blue line. By changing the temperature, magnetic field, and incident electron energy, the observed MR is consistent with theoretical expectations for spin-triplet MR. These results provide an important evidence for Rashba SOC induced spin-triplet SC at the $Fe_{0.29}TaS_2/Al_2O_3$/NbN interface. More importantly, such spin-triplet MR is theoretically calculated to be highly dependent on the normalized SOC ($\lambda$) and barrier ($\mathbf{Z}$) strengths [65], as shown in Fig. 4(d). It is noted that the nonmonotonic behavior of the interface barrier dependence is one of the key features for SOC-induced spin-triplet Andreev reflection. This feature is very different from the conventional Andreev reflection which monotonically decreases as barrier strength increases [63]. Other experimental signatures of SOC induced spin-triplet paring has also been observed in all-epitaxial Fe/MgO/V superconducting tunneling junction [66] and Nb/Pt/Co/Pt heterostructures[67]. The predicted anisotropy of the MR [64], which is enhanced by the presence of the spin-triplet Andreev reflections [62,65] has also important ramifications for the superconducting junctions with antiferromagnets where it could be used to probe the Neel vector direction and give unusual thermal transport properties [68].

Besides the electrical transport method including Josephson effect and spin-triplet Andreev reflection, another experimental method to probe the equal-spin-triplet Cooper pair takes advantage that they can carry spin angular momentum. For example, the spin-triplet Cooper pair mediated the dissipationless spin current in SC/FM heterostructures will induce a field-like toque and make the resonance field shift [69]. This has been experimentally investigated by Li et al. in the Nb/Py/Nb heterostructure via ferromagnetic resonance (FMR) method [70]. Subsequently, Jeon et al. reported that the Meissner screening effect can be another mechanism to induce the resonance filed shift in SC/FM heterostructures [71]. For the damping-like toque, Jeon et al. fabricated the Pt/Nb/Py/Nb/Pt heterostructures and observed the enhanced Gilbert damping below the superconducting temperature [72], which can be attributed to the spin-triplet supercurrent flowing across the Nb layer and relaxed in the heavy metal Pt with strong SOC. However, Silaev theoretically pointed out that Andreev bound states also can enhance the Gilbert damping in SC based heterostructures [73,74]. Therefore, definitive measurement of the spin-transfer torque needs further exploration.

In addition to the above two schemes based on real space and $\mathbf{K}$ space spin texture, Takahashi et al. and Houzet independently proposed that FMR can be a method for the generation of spin-triplet Cooper pairs due to the time-dependent spin texture [75,76].

The theoretical scheme provides a new perspective for the construction of spin-triplet SC where the transformation of spin quantization axis is realized by the spin texture in the time series. Remarkably, by controlling the resonant condition, an ultrafast turn on/off spin-triplet SC could be realized.

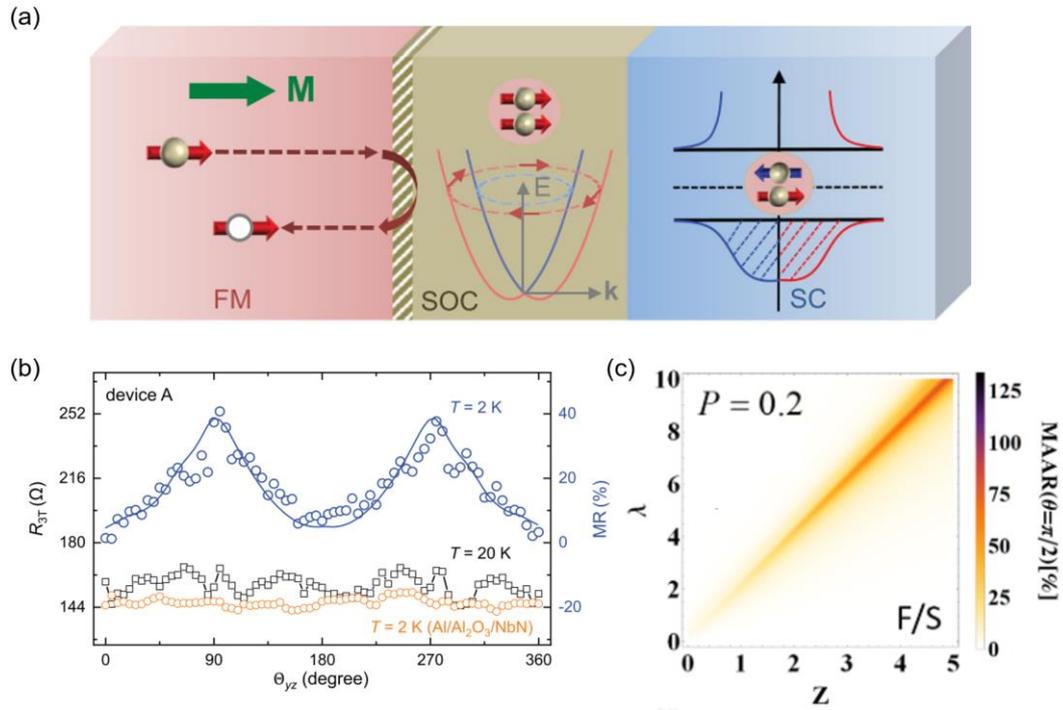

**Figure 4. Anisotropic spin-triplet SC and Andreev reflection at SC/FM interface due to Rashba SOC.** (a) The schematic of interfacial SOC induced the spin-triplet Cooper pair at the SC/FM interface. (b) Experimental observation of the two-fold anisotropic spin-triplet Andreev reflection magnetoresistance (named spin-triplet MR) at vdWs FM $Fe_{0.29}TaS_2$/SC heterostructures at $T = 2$ K (blue line). The two-fold MR behavior disappear at $T = 20$ K (black line) and in control device normal metal Al/SC heterostructure at $T = 2$ K (yellow line). (c) Theoretical simulation of the phase diagram of magneto-anisotropic Andreev reflection (MAAR) as a function of normalized SOC ($\lambda$) and barrier (**Z**) strengths. (a-b) are adapted from Ref [62], with permission, copyright Springer Nature 2021. (c) is adapted from Ref [65], with permission, copyright APS 2020.

## 2.4. 0-π transition in FM Josephson junction

In the FM/SC heterostructures, the superconducting parameter penetrating into the FM layer has an oscillatory decay that arises from the ferromagnetic exchange interaction-induced spin band imbalance in FM. The pairing at SC/FM interface gives rise to the center-of-mass momentum $\pm\hbar Q = \pm(\hbar k_{F\uparrow} - \hbar k_{F\downarrow})$ [10], where the $k_{F\uparrow}$ and $k_{F\downarrow}$ represent the Fermi vectors of majority and minority spin bands. Hence, the penetrating Cooper pair wave function oscillates with respect to the FM thickness, as illustrated in Fig. 5(a). For the SC/FM/SC heterostructures, the order parameters in the superconductors on both sides will have different phase differences (Δφ) depending on the thickness and the exchange interaction of FM layer. When $\Delta\varphi \in (0, \pi)$, positive Josephson coupling is formed, which is called 0-phase Josephson effect; When $\Delta\varphi = (\pi, 2\pi)$, it called π-phase Josephson coupling with negative Josephson coupling characteristics. The thickness tunable Josephson phase transition is a unique physical phenomenon arising from the quantum interplay between the ferromagnetism and superconductivity. Interestingly, another path to realize 0-π phase transition was predicted to employ Josephson junctions with metallic antiferromagnets [77].

The experimental observation of 0-π phase transition of FM Josephson junction via the Josephson critical current ($I_C$) oscillatory with FM thickness was first demonstrated in weak ferromagnets CuNi [78,79] and PdNi [80]. Apart from that, the temperature induced 0-π transition also has been confirmed in experiments which can be attributed to the temperature-dependent exchange coupling energy in a weak FM [78]. Subsequently, Robinson et al. [81] demonstrated a multi-oscillation behavior with the FM thickness in a strong FM Josephson junction including Co, Ni, and Py. As shown in the Fig. 5(b), the product of $I_C$ and normal state resistance ($R_N$) confirms multiple oscillation respect to Co thickness with the period ~ 1.9 nm. The experimental results can be simulated well by the theoretical model as illustrated by the solid line in Fig. 5(b). To directly recognize 0-π ground states, phase-sensitive measurements have been performed to investigate the current-phase relationships in the 0-π FM Josephson junctions [82-84]. These experimental evidences of the 0-π Josephson junction could pave the way for further superconducting logic devices and rapid-single-flux-quantum

(RSFQ) applications [85,86]. Towards these goals, an efficient manipulation of 0-π transition has been demonstrated theoretically [87,88] and experimentally [89,90] via varying relative orientation of the spin-valve structures embed in Josephson junction.

Besides the DC Josephson coupling effect, the dynamic spin properties of 0-π FM Josephson junctions have been recently investigated by Yao et al. [91] via FMR method. As shown in the Fig. 5(c), the Gilbert damping in the Nb/Py/Nb heterostructures reveals a giant oscillatory behavior as a function of the Py thickness. This observation suggests that different spin pumping and relaxation rates in 0-π FM Josephson junctions, which could be strongly affected by the energy levels of the Andreev bound states (ABS) [74,92-94]. The energy of ABS ($E_A$) is determined by the superconducting phase difference (Δφ) of the two SC leads [95,96]:

$$E_A = \pm \Delta_{sc} \sqrt{1 - D \sin^2 \frac{\Delta\varphi}{2}}, \qquad (3)$$

where D is the junction transparency. For π Josephson junction, the evanescent ABS quasiparticles group near the zero energy to dissipate the spin angular momentum, as shown in the inset of Fig. 5(c). On the other hand, for 0 Josephson junction, spin pumping via ABS is largely suppressed at low temperature since ABS reside near SC energy gap. As a result, the more efficient spin angular momentum relaxation in π Josephson junctions, the larger Gilbert damping is obtained. Therefore, the different spin relaxation in 0 and π Josephson junctions will give rise to the large oscillating Gilbert damping with respect to the FM thickness.

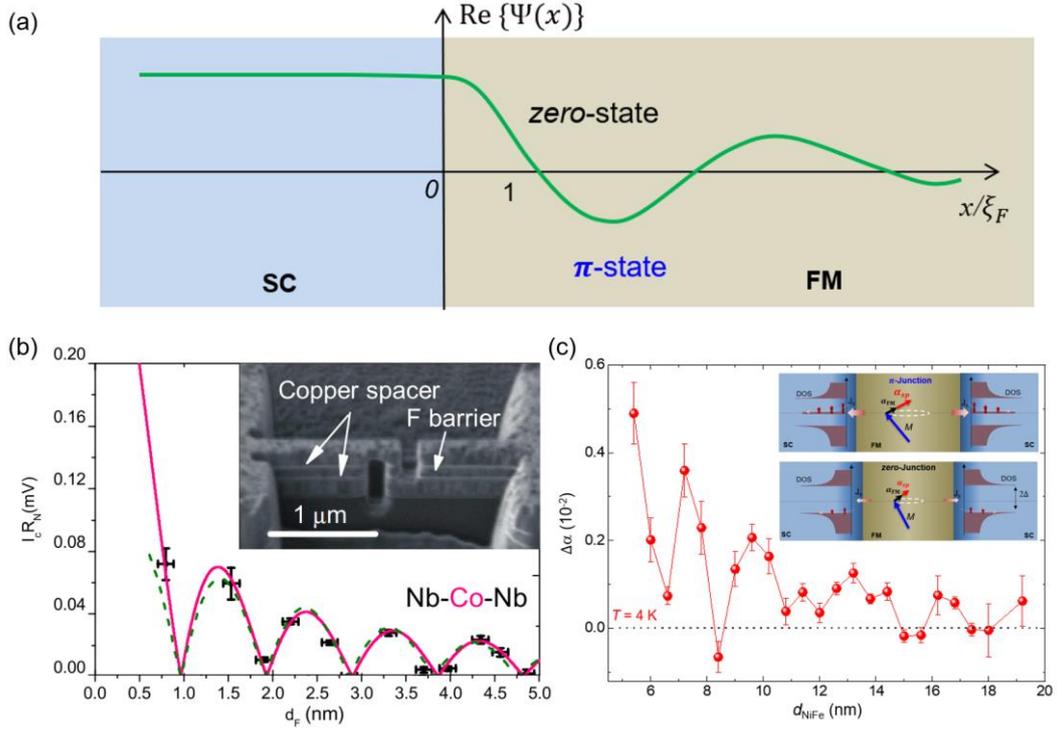

**Figure 5. Experimental progresses of 0-π FM Josephson junctions.** (a) The schematic of superconducting parameter oscillation with respect to the propagation distance in FM. (b) The oscillation of $I_cR_N$ (product of Josephson critical current and normal state resistance) as a function of FM thickness in Nb/Co/Nb junction at $T = 4.2$ K. Inset: The scanning electron microcopy of the Co Josephson junction devices fabricated using focused ion beam (FIB) etching technique. (c) The giant oscillatory behavior of spin dynamic parameter ($\Delta\alpha$) as a function of the FM thickness in Nb/Py/Nb junction at $T = 4$ K. Inset: physical pictures of the spin pumping via Andreev bound states of the π/0 Josephson junction. (a, c) are adapted from Ref [91], with permission, copyright American Association for the Advancement of Science 2021. (b) is adapted from Ref [81], with permission, copyright APS 2006.

## 3. SC/FM heterostructure for quantum computation
### 3.1. π Qubit

Superconducting quantum circuits can be constructed based on the Josephson junctions, which is considered as a promising route to realize quantum computers [97]. The three basic types of superconducting qubits [27] are the charge, flux, and phase qubits. One of their major challenges is the short coherence properties since they are sensitive to environmental charge and magnetic noise [12]. To overcome this obstacle, transmon [98], Xmon [99], fluxonium [100] are extensively explored in superconducting quantum computation with the cost of increasing circuit complexity and decreasing the integration ability [15,16]. Owing to the advances of π Josephson junction Yamashita et al. [12] theoretically proposed the concept of π flux qubit [101] using FM regions with two major advantages compared to conventional flux qubits: (i) it can operate at zero-magnetic field; thus, a long coherence time is expected. (ii) the small size of π flux qubit has the potential for a larger-scale integration. As shown in Fig. 6(a), the π flux qubit is constructed by a superconducting ring with an insulator Josephson junction and a π FM Josephson junction. In such a flux qubit, two degenerate potential minimums in superconducting phase space are formed: clockwise $|\uparrow\rangle$ and anti-clockwise $|\downarrow\rangle$ supercurrent states. Due to the quantum tunneling between $|\uparrow\rangle$ and $|\downarrow\rangle$, the bonding ($|0\rangle \propto |\uparrow\rangle + |\downarrow\rangle$) and anti-bonding ($|1\rangle \propto |\uparrow\rangle - |\downarrow\rangle$) states form the two-level quantum systems, which can be used to code for quantum computation. To explore the coherence properties for a π flux qubit, Kato et al. [14], estimated the relaxation time ($\tau_{relax}$) and pure dephasing time ($T_2^*$) based on the experimental parameters in FM Josephson junctions [102]. Figure 6(b) shows the $\tau_{relax}$ and $T_2^*$ of a π flux qubit as a function of the critical Josephson current density ($j_C$) for two sizes of $10 \times 10$ $\mu m^2$ (solid line) and $1 \times 1$ $\mu m^2$ (dash line). The coherence time can reach the order of 1 ms for a $10 \times 10$ $\mu m^2$ junction when $j_C$ up to $10^7$ A/$m^2$.

Besides the π flux qubit, π phase qubit has also been proposed by Noh et al., in which the information is directly coded on 0 and π phase Josephson coupling states [13]. Possible decoherence sources of π qubits include the spin-flip scattering [103] and dynamic response of magnetic domain structure in FM layer [104]. Experimentally, Feofanov et al. [86] embedded a π Josephson junction into a conventional phase qubit circuit and observed a clear Rabi oscillation [Fig. 6(c)]. Nearly identical decay time for the phase qubit with and without a π Josephson junction was observed, which indicates that adding a FM Josephson junction will not contribute additional dissipation in the superconducting quantum circuits.

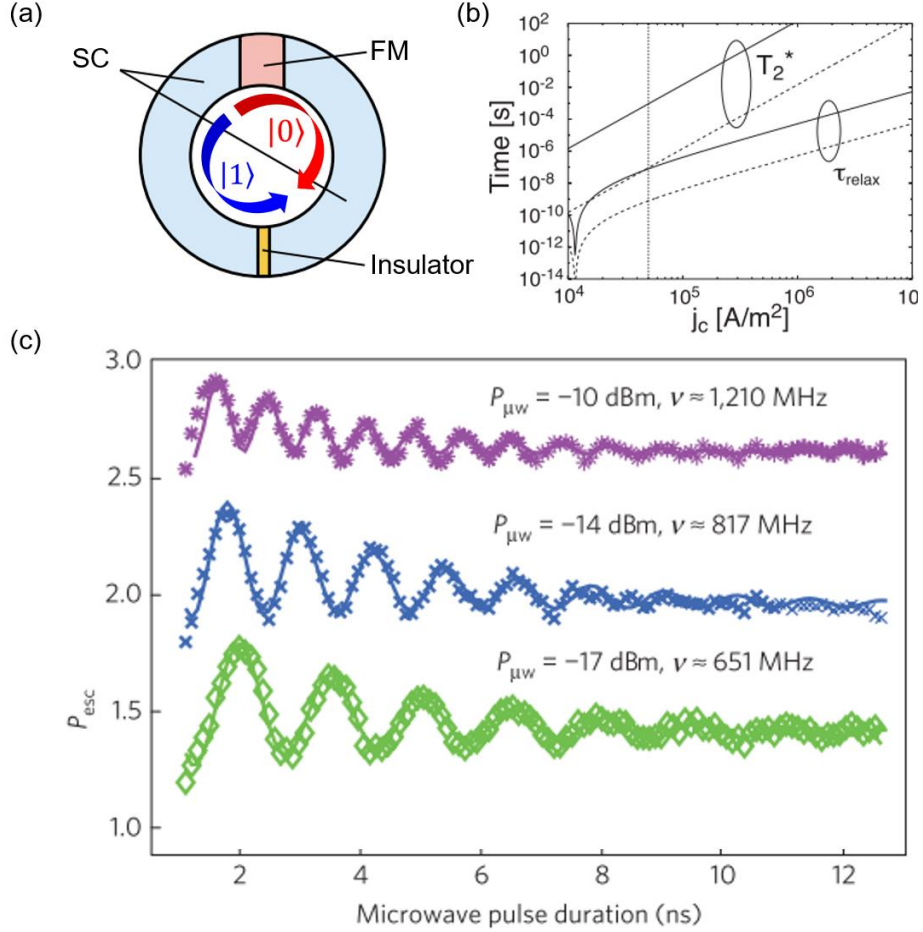

**Figure 6. π qubit.** (a) The schematic of π flux qubit consisting of a superconducting ring with an insulator (0 Josephson junction) and a FM spacer (π Josephson junction). (b) Critical Josephson current density ($j_C$) dependent relaxation time ($\tau_{relax}$) and pure dephasing time ($T_2^*$) for π flux qubits. The solid and dashed lines show the FM Josephson junction with area of $10 \times 10$ $\mu m^2$ and $1 \times 1$ $\mu m^2$, respectively. (c) Experimental Rabi oscillation results for conventional phase qubit embed with a π FM Josephson junction. (a) is adapted from Ref [12], with permission, copyright APS 2005. (b) is adapted from Ref [14], with permission, copyright APS 2007. (c) is adapted from Ref [86], with permission, copyright Springer Nature 2010.

### 3.2. Majorana states in SC/FM heterostructures

As a special real solution of the Dirac equation [105], Majorana fermion [18] is equal to its antiparticle. Unlike this original prediction in the context of a high-energy physics, the prospect of realizing such a peculiar behavior in condensed matter systems is even more fascinating as the underlying Majorana states are not really fermions and instead obey the non-Abelian statistics [106]. Due to the topological protection of the braiding processes, Majorana state are proposed to be ideal candidates for fault-tolerant quantum computation [107]. Various realization of Majorana states have been suggested by considering collective excitations in condensed matter systems. Specifically, Kitaev has proposed that the Majorana bound states (MBS) could form at the end of an one-dimensional (1D) spinless *p*-wave SC chain via tuning the nearest neighbor hopping strength and superconducting gap into topological SC phase [17]. A common experimental support of the MBS is the quantized zero bias conductance peak (ZBCP), first predicted in the studies of *p*-wave organic superconductors [108], cited also in Kitaev's work [17], as the materials implementation of MBS was initially unclear, while many additional ZBCP analyses and the role of disorder were subsequently considered [109]. By extending it to two-dimensional (2D), topological SC hosts Majorana chiral edge states with linear dispersion [110]. However, the observation of the desirable *p*-wave spin-triplet SC, which could support Majorana states remains elusive in the nature [43,44,111,112]. This caution of the claimed definitive *p*-wave spin-triplet signatures [113,114] is well illustrated in $Sr_2RuO_4$, suggested to also come from spin-singlet superconductivity [115], while recent experiments argue against spin-triplet superconductivity, even involving its discoverer, Y. Maeno [116,117].

Therefore, to realize the *p*-wave SC and Majorana states, Fu and Kane's seminal proposal [118] was very influential as it showed how they can be implemented using heterostructures and superconducting proximity effects. This idea has motivated many pioneering efforts devoted to artificial heterostructures, including strong SOC nanowire/SC [119-122], topological insulator/SC [118,123], and FM/SC [20,21,124-126] heterostructures. Here, we focus on the discussion of the FM/SC heterostructures, which could be used as promising platforms for the realization, manipulation, as well as braiding and fusion of Majorana states. This topic has been recently reviewed in Ref. [127]. Despite an impressive materials progress and intensive experimental effort, a conclusive demonstration of Majorana states is still missing.

To realize the MBS in FM/SC heterostructures, the key role is the interplay between the superconducting proximity effect and the spin textures, which is similar to the SOC induced spin-triplet SC [62,64]. One approach is theoretically proposed by depositing ferromagnetic metal chain onto the conventional *s*-wave superconducting thin films with strong SOC ($E_{so}$) [128]. The majority (spin-up) and minority (spin-down) Bogoliubov quasiparticle energy bands are shifted with the exchange splitting energy (*J*), as shown in the Fig. 7(a). Due to the strong SOC, the hybridization of Bogoliubov electron-like and hole-like quasiparticles in the spin-minority band make the superconducting pair potential (Δ) in FM chain reverses the spin character which results in the superconducting pairing between the same spin-polarized quasiparticles and the formation of the *p*-wave SC chain. Experimentally, Nadj-Perge et al. observed the ZBCP at the two ends of a ferromagnetic Fe atomic chain self-assembly formed on Pb superconducting substrate via the high-resolution scanning tunneling microscopy (STM) method [Fig. 7(b)] [20]. As shown in Fig. 7(c), the maximum ZBCP appears at the two ends of the Fe atomic chain, which decays rapidly in the range of 1 nm along the chain, interpreted to be compatible with MBS. The residual conductivity at the middle of the chain with small oscillatory behavior was attributed to the thermally broadened Shiba states. Experimental signatures of MBS have also been reported in 1D FM/SC heterostructures such as Co/Pb [126] and Fe/Re [125].

In 2D FM/SC heterostructures, Manna et al. fabricated the V/Au/EuS devices and observed the signature of Majorana zero mode (MZM) [23]. In this structure, the topological SC is formed in the Au (111) Shockley surface state with strong SOC up to 110 meV and proximitized from the superconducting V and ferromagnetic EuS [129,130]. The ZBCPs are reported at the edge of EuS islands under the in-plane magnetic field, which is consistent with theoretical prediction of MZM [131]. Recently, the signatures of chiral Majorana edge modes have also been reported in 2D vdWs heterostructures NbSe$_2$/CrBr$_3$ [22], as shown in Fig. 7(e).

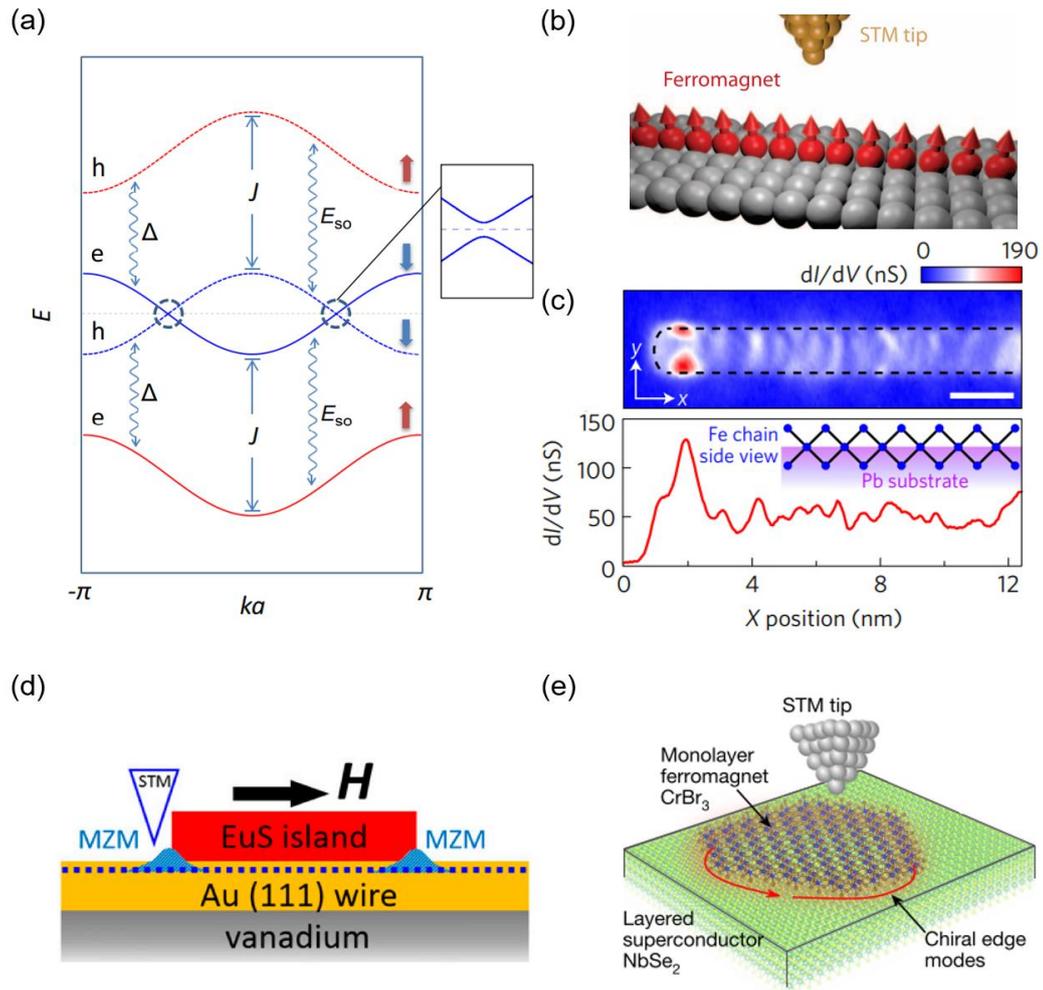

**Figure 7. Signatures of Majorana states in FM/SC heterostructures.** (a) The energy dispersion of Bogoliubov quasiparticles in a system with exchange splitting energy ($J$), superconducting pairing potential ($\Delta$) and SOC ($E_{so}$). The red (dash) and blue (dash) lines are corresponded to majority electron (hole) like quasiparticle band and minority electron (hole) like quasiparticle band, respectively. (b) The schematic of probing MBS at Fe atomic chain on Pb superconducting substrate. (c) The spatial dependence of zero-bias conductance (dI/dV) along the Fe atomic chain. (d) The schematic of Majorana zero modes (MZM) at V/Au/EuS heterostructures. (e) The schematic of the chiral Majorana edge modes at 2D vdWs $NbSe_2$/$CrBr_3$ heterostructures. (a) is adapted from Ref [128], with permission, copyright APS 2021. (b) is adapted from Ref [20], with permission, copyright American Association for the Advancement of Science 2014. (c) is adapted from Ref [124], with permission, copyright Springer Nature 2017. (d) is adapted from Ref [23], with permission, copyright Springer PNAS 2020. (e) is adapted from Ref [22], with permission, copyright Springer Nature 2020.

The braiding and, experimentally simpler, fusion of Majorana states have been considered as the key step for topological quantum computing [19,132-134]. They can serve to implement quantum gates as well as to probe the non-Abelian statistics and overcome various challenges and spurious effects in experimentally confirming Majorana states using spectral features, such as ZBCP [135]. While many proposals for braiding and fusion have been known for a while, especially in 1D nonmagnetic systems [19,132], so far, they have not been experimentally demonstrated. Even for a simpler fusion, in extensively studied 1D semiconductor nanowires [120,121], one can identify a number of obstacles, from fine-tuned parameters required for topological superconductivity and constraints of 1D geometry, to a missing accurate preparation of the initial state [136]. This situation serves as a caution in the effort to bridge the gap from theoretical proposals for braiding and fusion of MBS in FM/SC heterostructures and their experimental implementation. For example, manipulating different magnetic structures, from atomic chains [137] and domain walls [138] to skyrmions [139]. Among them, Li et al. [137], who proposed that a magnetic field can be used to manipulate the topological phase transition of helix magnetization atomic chain to efficient control of MBS [Fig. 8(a)] in the 1D SC/FM heterostructures.

An alternative approach for MBS braiding, brings together advances in spintronics, proximity effects, and topological superconductivity to recognize that it is important to use 2D platforms [140]. It was recognized that electrically-tunable magnetic textures could provide synthetic SOC (even in the absence of native SOC), Zeeman splitting, and confinement to create, control, and braid in the superconducting proximitized two-dimensional electron gas (2DEG) [140]. Instead of arrays of magnetic tunnel junctions [140], subsequent studies have considered a more feasible approach relying on commercially available spin valves [141]. However, unlike in spintronic applications, where the focus is on magnetoresistive effects, here the functionality of spin valves pertains to their tunable stray fields. As shown in the Fig. 8(b), the spatially-dependent stray fields from tunable magnetic textures can be used to induce the topological SC in different regions of 2DEG. One advantage of these MBS is that they can be efficiently manipulated via spin-transfer torque-controlled fringing fields of the spin valves [142-

144]. A similar method of using magnetic stripes [145] has also been theoretically proposed. One advantage of these approaches is that the spintronics technology is quite mature and could facilitate integrating the 2D SC/FM heterostructures for topological quantum computing. Experimentally, the feasibility of a similar scheme is supported by the generation of sufficiently strong synthetic SOC, which is reported to support MBS in proximitized carbon nanotubes [146].

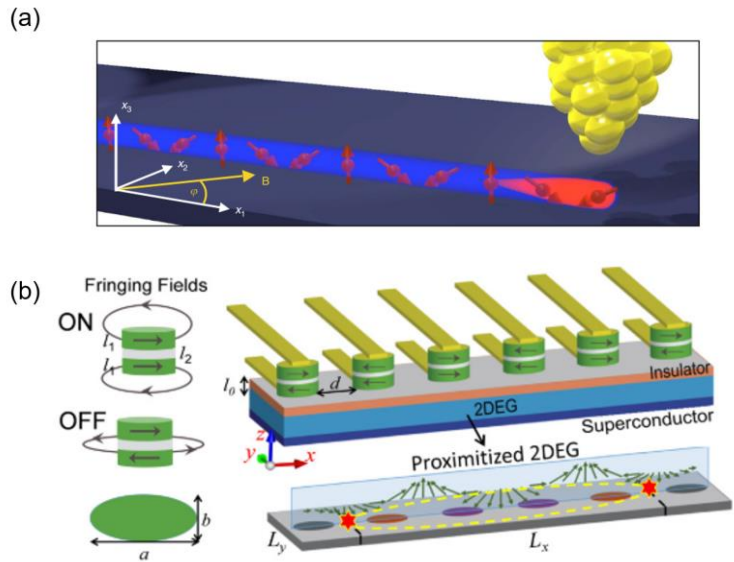

**Figure 8. Theoretical proposals of manipulating MBS in the SC/FM heterostructures.** (a) The schematic of manipulating MBS via external magnetic field in an atomic chain with a helical magnetization, self-assembled on a superconducting substrate. (b) The schematic of MBS manipulation of using the commercial spin-valve arrays on superconducting proximitized two-dimensional electron gas (2DEG) system. (a) is adapted from Ref [137], with permission, copyright Springer Nature 2016. (b) is adapted from Ref [141], with permission, copyright APS 2019.

## 4. Summary and outlook

The novel physical properties and materials advances in SC/FM heterostructure have greatly promoted a rapid progress of superconducting spintronics and quantum computing. Looking forward, we expect a direct connection of experimental investigation of the spin-triplet SC and *p*-wave SC in SC/FM heterostructures by optimizing device structures to attain ballistic transport [147] or using the topological

FM [148-150]. Furthermore, the interaction of superconductivity, ferromagnetism, and topology might also give rise to additional novel phenomena, such as the Josephson diode effect [151,152], which was also previously seen in the studies of Al/InAs 2DEG junctions that could support topological superconductivity [153,154] and can have interesting time-dependent manifestations [155]. Furthemore, vortex diode effect has also been explored in SC/FM heterostructures very recently [156], which shows the potential to realize the on-chip microwave filter for future superconducting quantum circuit application. The realization of FMR induced spin-triplet SC [75,76] will be not only a new experimental strategy to realize spin-triplet SC, but also a foundation for the coupling between microwave photon and SC/FM systems. The strong coupling might lay the foundation of further manipulation of π qubits based on SC/FM heterostructures. To embed SC/FM heterostructures into practical superconducting quantum circuits, optimizing their quality and interfacial properties will be the key steps.

## Acknowledgements

We acknowledge the financial support from National Basic Research Programs of China (Nos. 2019YFA0308401), National Natural Science Foundation of China (Nos. 11974025), and the Key Research Program of the Chinese Academy of Sciences (Grant No. XDB28000000), U.S. NSF ECCS-2130845, U.S. ONR N000141712793, U.S. DOE, Office of Science BES, Award No. DE-SC0004890, and U.S. ONR MURI N000142212764.

# Reference

[1]   J. Bardeen, L. N. Cooper, and J. R. Schrieffer, Microscopic Theory of Superconductivity. *Physical Review* **106**, 162 (1957).

[2]   P. Fulde and R. A. Ferrell, Superconductivity in a Strong Spin-Exchange Field. *Physical Review* **135**, A550 (1964).

[3]   A. I. Larkin and Y. N. Ovchinnikov, Nonuniform State of Superconductors. *Soviet Physics-JETP* **20**, 762 (1965).

[4]   A. I. Buzdin, Proximity effects in superconductor-ferromagnet heterostructures. *Reviews of Modern Physics* **77**, 935 (2005).

[5]   B. Li, N. Roschewsky, B. A. Assaf, M. Eich, M. Epstein-Martin, D. Heiman, M. Münzenberg, and J. S. Moodera, Superconducting Spin Switch with Infinite Magnetoresistance Induced by an Internal Exchange Field. *Physical Review Letters* **110**, 097001 (2013).

[6]   W. Han, R. K. Kawakami, M. Gmitra, and J. Fabian, Graphene spintronics. *Nat Nanotechnol* **9**, 794 (2014).

[7]   I. Žutić, J. Fabian, and S. Das Sarma, Spintronics: Fundamentals and applications. *Reviews of Modern Physics* **76**, 323 (2004).

[8]   H. Yang, S.-H. Yang, S. Takahashi, S. Maekawa, and S. S. P. Parkin, Extremely long quasiparticle spin lifetimes in superconducting aluminium using MgO tunnel spin injectors. *Nature Materials* **9**, 586 (2010).

[9]   S. K. Kim, R. Myers, and Y. Tserkovnyak, Nonlocal Spin Transport Mediated by a Vortex Liquid in Superconductors. *Physical Review Letters* **121**, 187203 (2018).

[10] M. Eschrig, Spin-polarized supercurrents for spintronics: a review of current progress. *Reports on Progress in Physics* **78**, 104501 (2015).

[11] J. Linder and J. W. A. Robinson, Superconducting spintronics. *Nature Physics* **11**, 307 (2015).

[12] T. Yamashita, K. Tanikawa, S. Takahashi, and S. Maekawa, Superconducting π Qubit with a Ferromagnetic Josephson Junction. *Physical Review Letters* **95**, 097001 (2005).

[13] T.-W. Noh, H. S. Sim, and M. D. Kim, Superconductor-ferromagnet-junction phase qubit. *Journal of the Korean Physical Society* **60**, 72 (2012).

[14] T. Kato, A. A. Golubov, and Y. Nakamura, Decoherence in a superconducting flux qubit with a π-junction. *Physical Review B* **76**, 172502 (2007).

[15] G.-L. Guo, H.-B. Leng, Y. Hu, and X. Liu, 0-π qubit with one Josephson junction. *Physical Review B* **105**, L180502 (2022).

[16] J. B. Trebbia, Q. Deplano, P. Tamarat, and B. Lounis, Tailoring the superradiant and subradiant nature of two coherently coupled quantum emitters. *Nature Communications* **13**, 2962 (2022).

[17] A. Y. Kitaev, Unpaired Majorana fermions in quantum wires. *Physics-Uspekhi* **44**, 131 (2001).

[18] E. Majorana, Teoria simmetrica dell'elettrone e del positrone. *Il Nuovo Cimento* **14**, 171 (1937).

[19] D. Aasen, M. Hell, R. V. Mishmash, A. Higginbotham, J. Danon, M. Leijnse, T. S. Jespersen, J. A. Folk, C. M. Marcus, K. Flensberg, and J. Alicea, Milestones Toward Majorana-Based Quantum Computing. *Physical Review X* **6**, 031016 (2016).

[20] S. Nadj-Perge, K. Drozdov Ilya, J. Li, H. Chen, S. Jeon, J. Seo, H. MacDonald Allan, B. A. Bernevig, and A. Yazdani, Observation of Majorana fermions in ferromagnetic atomic chains on a superconductor. *Science* **346**, 602 (2014).

[21] B. Jäck, Y. Xie, J. Li, S. Jeon, B. A. Bernevig, and A. Yazdani, Observation of a Majorana zero mode in a topologically protected edge channel. *Science* **364**, 1255 (2019).

[22] S. Kezilebieke, M. N. Huda, V. Vaňo, M. Aapro, S. C. Ganguli, O. J. Silveira, S. Głodzik, A. S. Foster, T. Ojanen, and P. Liljeroth, Topological superconductivity in a van der Waals heterostructure. *Nature* **588**, 424 (2020).

[23] S. Manna, P. Wei, Y. Xie, T. Law Kam, A. Lee Patrick, and S. Moodera Jagadeesh, Signature of a pair of Majorana zero modes in superconducting gold surface states. *Proceedings of the National Academy of Sciences* **117**, 8775 (2020).

[24] B. Niedzielski and J. Berakdar, Controlled Vortex Formation at Nanostructured Superconductor/Ferromagnetic Junctions. *physica status solidi (b)* **257**, 1900709 (2020).

[25] W. Han, S. Maekawa, and X.-C. Xie, Spin current as a probe of quantum materials. *Nature Materials* **19**, 139 (2020).

[26] M. Eschrig, Spin-polarized supercurrents for spintronics. *Physics Today* **64**, 43 (2010).

[27] J. Clarke and F. K. Wilhelm, Superconducting quantum bits. *Nature* **453**, 1031 (2008).

[28] T. Wakamura, H. Akaike, Y. Omori, Y. Niimi, S. Takahashi, A. Fujimaki, S. Maekawa, and Y. Otani, Quasiparticle-mediated spin Hall effect in a superconductor. *Nat Mater* **14**, 675 (2015).

[29] C. W. J. Beenakker, Annihilation of Colliding Bogoliubov Quasiparticles Reveals their Majorana Nature. *Physical Review Letters* **112**, 070604 (2014).

[30] S. A. Kivelson and D. S. Rokhsar, Bogoliubov quasiparticles, spinons, and spin-charge decoupling in superconductors. *Physical Review B* **41**, 11693 (1990).

[31] B. Leridon, J. Lesueur, and M. Aprili, Spin-bottleneck due to spin-charge separation in a superconductor. *Physical Review B* **72**, 180505 (2005).

[32] T. Yamashita, S. Takahashi, H. Imamura, and S. Maekawa, Spin transport and relaxation in superconductors. *Physical Review B* **65**, 172509 (2002).

[33] C. H. L. Quay, D. Chevallier, C. Bena, and M. Aprili, Spin imbalance and spin-charge separation in a mesoscopic superconductor. *Nature Physics* **9**, 84 (2013).

[34] D. Gershoni, Quantum information: long live the spin. *Nat Mater* **5**, 255 (2006).

[35] Y. Yao, Q. Song, Y. Takamura, J. P. Cascales, W. Yuan, Y. Ma, Y. Yun, X. C. Xie, J. S. Moodera, and W. Han, Probe of spin dynamics in superconducting NbN thin films via spin pumping. *Physical Review B* **97**, 224414 (2018).

[36] M. Inoue, M. Ichioka, and H. Adachi, Spin pumping into superconductors: A new probe of spin dynamics in a superconducting thin film. *Physical Review B* **96**, 024414 (2017).

[37] A. J. Berger, E. R. J. Edwards, H. T. Nembach, A. D. Karenowska, M. Weiler, and T. J. Silva, Inductive detection of fieldlike and dampinglike ac inverse spin-orbit torques in ferromagnet/normal-metal bilayers. *Physical Review B* **97**, 094407 (2018).

[38] M. Müller, L. Liensberger, L. Flacke, H. Huebl, A. Kamra, W. Belzig, R. Gross, M. Weiler, and M. Althammer, Temperature-Dependent Spin Transport and Current-Induced Torques in Superconductor-Ferromagnet Heterostructures. *Physical Review Letters* **126**, 087201 (2021).

[39] A. A. Abrikosov, On the Magnetic Properties of Superconductors of the Second Group. **5**, 1174 (1957).

[40] L. V. Shubnikov, Khotkevich, V. I., Shepelev, Y. D. & Ryabinin, Y. N., Magnetic properties of superconductors and alloys. *Zh. Eksper. Teor. Fiz.* **7**, 221 (1937).

[41] O. V. Dobrovolskiy, R. Sachser, T. Brächer, T. Böttcher, V. V. Kruglyak, R. V. Vovk, V. A. Shklovskij, M. Huth, B. Hillebrands, and A. V. Chumak, Magnon–fluxon interaction in a ferromagnet/superconductor heterostructure. *Nature Physics* **15**, 477 (2019).

[42] V. Vlaminck and M. Bailleul, Current-Induced Spin-Wave Doppler Shift. *Science* **322**, 410 (2008).

[43] S. Ran, C. Eckberg, Q. P. Ding, Y. Furukawa, T. Metz, S. R. Saha, I. L. Liu, M. Zic, H. Kim, J. Paglione, and N. P. Butch, Nearly ferromagnetic spin-triplet superconductivity. *Science* **365**, 684 (2019).

[44] S. Ran, I. L. Liu, Y. S. Eo, D. J. Campbell, P. M. Neves, W. T. Fuhrman, S. R. Saha, C. Eckberg, H. Kim, D. Graf, F. Balakirev, J. Singleton, J. Paglione, and N. P. Butch, Extreme magnetic field-boosted superconductivity. *Nature Physics* **15**, 1250 (2019).

[45] Y. Cao, J. M. Park, K. Watanabe, T. Taniguchi, and P. Jarillo-Herrero, Pauli-limit violation and re-entrant superconductivity in moiré graphene. *Nature* **595**, 526 (2021).

[46] R. S. Keizer, S. T. B. Goennenwein, T. M. Klapwijk, G. Miao, G. Xiao, and A. Gupta, A spin triplet supercurrent through the half-metallic ferromagnet $CrO_2$. *Nature* **439**, 825 (2006).

[47] C. Visani, Z. Sefrioui, J. Tornos, C. Leon, J. Briatico, M. Bibes, A. Barthélémy, J. Santamaría, and J. E. Villegas, Equal-spin Andreev reflection and long-range coherent transport in high-temperature superconductor/half-metallic ferromagnet junctions. *Nature Physics* **8**, 539 (2012).

[48] D. Sanchez-Manzano, S. Mesoraca, F. A. Cuellar, M. Cabero, V. Rouco, G. Orfila, X. Palermo, A. Balan, L. Marcano, A. Sander, M. Rocci, J. Garcia-Barriocanal, F. Gallego, J. Tornos, A. Rivera, F. Mompean, M. Garcia-Hernandez, J. M. Gonzalez-Calbet, C. Leon, S. Valencia, C. Feuillet-Palma, N. Bergeal, A. I. Buzdin, J. Lesueur, J. E. Villegas, and J. Santamaria, Extremely long-range, high-temperature Josephson coupling across a half-metallic ferromagnet. *Nature Materials* **21**, 188 (2022).

[49] I. Žutić and S. Das Sarma, Spin-polarized transport and Andreev reflection in semiconductor/superconductor hybrid structures. *Physical Review B* **60**, R16322 (1999).

[50] F. S. Bergeret, A. F. Volkov, and K. B. Efetov, Long-Range Proximity Effects in Superconductor-Ferromagnet Structures. *Physical Review Letters* **86**, 4096 (2001).

[51] T. S. Khaire, M. A. Khasawneh, W. P. Pratt, and N. O. Birge, Observation of Spin-Triplet Superconductivity in Co-Based Josephson Junctions. *Physical Review Letters* **104**, 137002 (2010).

[52] J. W. A. Robinson, J. D. S. Witt, and M. G. Blamire, Controlled Injection of Spin-Triplet Supercurrents into a Strong Ferromagnet. *Science* **329**, 59 (2010).

[53] K.-R. Jeon, B. K. Hazra, K. Cho, A. Chakraborty, J.-C. Jeon, H. Han, H. L. Meyerheim, T. Kontos, and S. S. P. Parkin, Long-range supercurrents through a chiral non-collinear antiferromagnet in lateral Josephson junctions. *Nature Materials* **20**, 1358 (2021).

[54] M. S. Anwar, F. Czeschka, M. Hesselberth, M. Porcu, and J. Aarts, Long-range supercurrents through half-metallic ferromagnetic $CrO_2$. *Physical Review B* **82**, 100501 (2010).

[55] A. Singh, C. Jansen, K. Lahabi, and J. Aarts, High-Quality $CrO_2$ Nanowires for Dissipation-less Spintronics. *Physical Review X* **6**, 041012 (2016).

[56] M. Eschrig and T. Löfwander, Triplet supercurrents in clean and disordered half-metallic ferromagnets. *Nature Physics* **4**, 138 (2008).

[57] M. S. Anwar, M. Veldhorst, A. Brinkman, and J. Aarts, Long range supercurrents in ferromagnetic $CrO2$ using a multilayer contact structure. *Applied Physics Letters* **100**, 052602 (2012).

[58] A. Singh, S. Voltan, K. Lahabi, and J. Aarts, Colossal Proximity Effect in a Superconducting Triplet Spin Valve Based on the Half-Metallic Ferromagnet $CrO_2$. *Physical Review X* **5**, 021019 (2015).

[59] N. Banerjee, J. W. A. Robinson, and M. G. Blamire, Reversible control of spin-polarized supercurrents in ferromagnetic Josephson junctions. *Nature Communications* **5**, 4771 (2014).

[60] J. W. A. Robinson, F. Chiodi, M. Egilmez, G. B. Halász, and M. G. Blamire, Supercurrent enhancement in Bloch domain walls. *Scientific Reports* **2**, 699 (2012).

[61] L. P. Gor'kov and E. I. Rashba, Superconducting 2D System with Lifted Spin Degeneracy: Mixed Singlet-Triplet State. *Physical Review Letters* **87**, 037004 (2001).

[62] R. Cai, Y. Yao, P. Lv, Y. Ma, W. Xing, B. Li, Y. Ji, H. Zhou, C. Shen, S. Jia, X. C. Xie, I. Žutić, Q.-F. Sun, and W. Han, Evidence for anisotropic spin-triplet Andreev reflection at the 2D van der Waals ferromagnet/superconductor interface. *Nature Communications* **12**, 6725 (2021).

[63] G. E. Blonder, M. Tinkham, and T. M. Klapwijk, Transition from metallic to tunneling regimes in superconducting microconstrictions: Excess current, charge imbalance, and supercurrent conversion. *Physical Review B* **25**, 4515 (1982).

[64] P. Högl, A. Matos-Abiague, I. Žutić, and J. Fabian, Magnetoanisotropic Andreev Reflection in Ferromagnet-Superconductor Junctions. *Physical Review Letters* **115**, 116601 (2015).

[65] T. Vezin, C. Shen, J. E. Han, and I. Žutić, Enhanced spin-triplet pairing in magnetic junctions with s-wave superconductors. *Physical Review B* **101**, 014515 (2020).

[66] I. Martínez, P. Högl, C. González-Ruano, J. P. Cascales, C. Tiusan, Y. Lu, M. Hehn, A. Matos-Abiague, J. Fabian, I. Žutić, and F. G. Aliev, Interfacial Spin-Orbit Coupling: A Platform for Superconducting Spintronics. *Physical Review Applied* **13**, 014030 (2020).

[67] N. Banerjee, J. A. Ouassou, Y. Zhu, N. A. Stelmashenko, J. Linder, and M. G. Blamire, Controlling the superconducting transition by spin-orbit coupling. *Physical Review B* **97**, 184521 (2018).

[68] M. F. Jakobsen, K. B. Naess, P. Dutta, A. Brataas, and A. Qaiumzadeh, Electrical and thermal transport in antiferromagnet-superconductor junctions. *Physical Review B* **102**, 140504 (2020).

[69] C. Holmqvist, S. Teber, and M. Fogelström, Nonequilibrium effects in a Josephson junction coupled to a precessing spin. *Physical Review B* **83**, 104521 (2011).

[70] L.-L. Li, Y.-L. Zhao, X.-X. Zhang, and Y. Sun, Possible Evidence for Spin-Transfer Torque Induced by Spin-Triplet Supercurrents. *Chinese Physics Letters* **35**, 077401 (2018).

[71] K.-R. Jeon, C. Ciccarelli, H. Kurebayashi, L. F. Cohen, X. Montiel, M. Eschrig, T. Wagner, S. Komori, A. Srivastava, J. W. A. Robinson, and M. G. Blamire, Effect of Meissner Screening and Trapped Magnetic Flux on Magnetization Dynamics in Thick Nb/Ni80Fe20/Nb Trilayers. *Physical Review Applied* **11**, 014061 (2019).

[72] K.-R. Jeon, C. Ciccarelli, A. J. Ferguson, H. Kurebayashi, L. F. Cohen, X. Montiel, M. Eschrig, J. W. A. Robinson, and M. G. Blamire, Enhanced spin pumping into superconductors provides evidence for superconducting pure spin currents. *Nature Materials* **17**, 499 (2018).

[73] M. A. Silaev, Large enhancement of spin pumping due to the surface bound states in normal metal--superconductor structures. *Physical Review B* **102**, 180502 (2020).

[74] M. Tanhayi Ahari and Y. Tserkovnyak, Superconductivity-enhanced spin pumping: Role of Andreev resonances. *Physical Review B* **103**, L100406 (2021).

[75] S. Takahashi, S. Hikino, M. Mori, J. Martinek, and S. Maekawa, Supercurrent Pumping in Josephson Junctions with a Half-Metallic Ferromagnet. *Physical Review Letters* **99**, 057003 (2007).

[76] M. Houzet, Ferromagnetic Josephson Junction with Precessing Magnetization. *Physical Review Letters* **101**, 057009 (2008).

[77] L. Bulaevskii, R. Eneias, and A. Ferraz, Superconductor-antiferromagnet-superconductor π Josephson junction based on an antiferromagnetic barrier. *Physical Review B* **95**, 104513 (2017).

[78] V. V. Ryazanov, V. A. Oboznov, A. Y. Rusanov, A. V. Veretennikov, A. A. Golubov, and J. Aarts, Coupling of Two Superconductors through a Ferromagnet: Evidence for a π Junction. *Physical Review Letters* **86**, 2427 (2001).

[79] V. A. Oboznov, V. V. Bol'ginov, A. K. Feofanov, V. V. Ryazanov, and A. I. Buzdin, Thickness Dependence of the Josephson Ground States of Superconductor-Ferromagnet-Superconductor Junctions. *Physical Review Letters* **96**, 197003 (2006).

[80] T. Kontos, M. Aprili, J. Lesueur, F. Genêt, B. Stephanidis, and R. Boursier, Josephson Junction through a Thin Ferromagnetic Layer: Negative Coupling. *Physical Review Letters* **89**, 137007 (2002).

[81] J. W. A. Robinson, S. Piano, G. Burnell, C. Bell, and M. G. Blamire, Critical Current Oscillations in Strong Ferromagnetic π Junctions. *Physical Review Letters* **97**, 177003 (2006).

[82] M. A. Khasawneh, W. P. Pratt, and N. O. Birge, Josephson junctions with a synthetic antiferromagnetic interlayer. *Physical Review B* **80**, 020506 (2009).

[83] S. M. Frolov, D. J. Van Harlingen, V. A. Oboznov, V. V. Bolginov, and V. V. Ryazanov, Measurement of the current-phase relation of superconductor/ferromagnet/superconductor π Josephson junctions. *Physical Review B* **70**, 144505 (2004).

[84] W. Guichard, M. Aprili, O. Bourgeois, T. Kontos, J. Lesueur, and P. Gandit, Phase Sensitive Experiments in Ferromagnetic-Based Josephson Junctions. *Physical Review Letters* **90**, 167001 (2003).

[85] M. I. Khabipov, D. V. Balashov, F. Maibaum, A. B. Zorin, V. A. Oboznov, V. V. Bolginov, A. N. Rossolenko, and V. V. Ryazanov, A single flux quantum circuit with a ferromagnet-based Josephson π-junction. *Superconductor Science and Technology* **23**, 045032 (2010).

[86] A. K. Feofanov, V. A. Oboznov, V. V. Bol'ginov, J. Lisenfeld, S. Poletto, V. V. Ryazanov, A. N. Rossolenko, M. Khabipov, D. Balashov, A. B. Zorin, P. N. Dmitriev, V. P. Koshelets, and A. V. Ustinov, Implementation of superconductor/ferromagnet/ superconductor π-shifters in superconducting digital and quantum circuits. *Nature Physics* **6**, 593 (2010).

[87] Z. Pajović, M. Božović, Z. Radović, J. Cayssol, and A. Buzdin, Josephson coupling through ferromagnetic heterojunctions with noncollinear magnetizations. *Physical Review B* **74**, 184509 (2006).

[88] B. Crouzy, S. Tollis, and D. A. Ivanov, Josephson current in a superconductor-ferromagnet junction with two noncollinear magnetic domains. *Physical Review B* **75**, 054503 (2007).

[89] E. C. Gingrich, B. M. Niedzielski, J. A. Glick, Y. Wang, D. L. Miller, R. Loloee, W. P. Pratt Jr, and N. O. Birge, Controllable 0–π Josephson junctions containing a ferromagnetic spin valve. *Nature Physics* **12**, 564 (2016).

[90] J. A. Glick, V. Aguilar, A. B. Gougam, B. M. Niedzielski, E. C. Gingrich, L. Reza, W. P. Pratt, and N. O. Birge, Phase control in a spin-triplet SQUID. *Science Advances* **4**, eaat9457.

[91] Y. Yao, R. Cai, T. Yu, Y. Ma, W. Xing, Y. Ji, X.-C. Xie, S.-H. Yang, and W. Han, Giant oscillatory Gilbert damping in superconductor/ferromagnet/superconductor junctions. *Science Advances* **7**, eabh3686 (2021).

[92] A. A. Golubov, M. Y. Kupriyanov, and E. Il'ichev, The current-phase relation in Josephson junctions. *Reviews of Modern Physics* **76**, 411 (2004).

[93] Z. Y. Chen, A. Biswas, I. Žutić, T. Wu, S. B. Ogale, R. L. Greene, and T. Venkatesan, Spin-polarized transport across a $La_{0.7}Sr_{0.3}MnO_3/YBa_2Cu_3O_{7-x}$ interface: Role of Andreev bound states. *Physical Review B* **63**, 212508 (2001).

[94] S. Kashiwaya and Y. Tanaka, Tunnelling effects on surface bound states in unconventional superconductors. *Reports on Progress in Physics* **63**, 1641 (2000).

[95] F. Giazotto, J. T. Peltonen, M. Meschke, and J. P. Pekola, Superconducting quantum interference proximity transistor. *Nature Physics* **6**, 254 (2010).

[96] L. Bretheau, J. I. J. Wang, R. Pisoni, K. Watanabe, T. Taniguchi, and P. Jarillo-Herrero, Tunnelling spectroscopy of Andreev states in graphene. *Nature Physics* **13**, 756 (2017).

[97] H.-L. Huang, D. Wu, D. Fan, and X. Zhu, Superconducting quantum computing: a review. *Science China Information Sciences* **63**, 180501 (2020).

[98] J. Koch, T. M. Yu, J. Gambetta, A. A. Houck, D. I. Schuster, J. Majer, A. Blais, M. H. Devoret, S. M. Girvin, and R. J. Schoelkopf, Charge-insensitive qubit design derived from the Cooper pair box. *Physical Review A* **76**, 042319 (2007).

[99] R. Barends, J. Kelly, A. Megrant, D. Sank, E. Jeffrey, Y. Chen, Y. Yin, B. Chiaro, J. Mutus, C. Neill, P. O'Malley, P. Roushan, J. Wenner, T. C. White, A. N. Cleland, and J. M. Martinis, Coherent Josephson Qubit Suitable for Scalable Quantum Integrated Circuits. *Physical Review Letters* **111**, 080502 (2013).

[100] F. Bao, H. Deng, D. Ding, R. Gao, X. Gao, C. Huang, X. Jiang, H.-S. Ku, Z. Li, X. Ma, X. Ni, J. Qin, Z. Song, H. Sun, C. Tang, T. Wang, F. Wu, T. Xia, W. Yu, F. Zhang, G. Zhang, X. Zhang, J. Zhou, X. Zhu, Y. Shi, J. Chen, H.-H. Zhao, and C. Deng, Fluxonium: An Alternative Qubit Platform for High-Fidelity Operations. *Physical Review Letters* **129**, 010502 (2022).

[101] L. B. Ioffe, V. B. Geshkenbein, M. V. Feigel'man, A. L. Fauchère, and G. Blatter, Environmentally decoupled sds-wave Josephson junctions for quantum computing. *Nature* **398**, 679 (1999).

[102] M. Weides, M. Kemmler, E. Goldobin, D. Koelle, R. Kleiner, H. Kohlstedt, and A. Buzdin, High quality ferromagnetic 0 and π Josephson tunnel junctions. *Applied Physics Letters* **89**, 122511 (2006).

[103] R. Mélin, sin (2φ) current-phase relation in SFS junctions with decoherence in the ferromagnet. *Europhysics Letters (EPL)* **69**, 121 (2005).

[104] S.-i. Hikino, M. Mori, S. Takahashi, and S. Maekawa, Ferromagnetic Resonance Induced Josephson Current in a Superconductor/Ferromagnet/Superconductor Junction. *Journal of the Physical Society of Japan* **77**, 053707 (2008).

[105] P. A. M. Dirac, Quantum Mechanics of Many-Electron Systems. *Proceedings of the Royal Society of London Series A* **123**, 714 (1929).

[106] C. Nayak, S. H. Simon, A. Stern, M. Freedman, and S. Das Sarma, Non-Abelian anyons and topological quantum computation. *Reviews of Modern Physics* **80**, 1083 (2008).

[107] A. Y. Kitaev, Fault-tolerant quantum computation by anyons. *Annals of Physics* **303**, 2 (2003).

[108]    K. Sengupta, I. Žutić, H.-J. Kwon, V. M. Yakovenko, and S. Das Sarma, Midgap edge states and pairing symmetry of quasi-one-dimensional organic superconductors. *Physical Review B* **63**, 144531 (2001).

[109]    K. T. Law, P. A. Lee, and T. K. Ng, Majorana Fermion Induced Resonant Andreev Reflection. *Physical Review Letters* **103**, 237001 (2009).

[110]    M. Z. Hasan and C. L. Kane, Colloquium: Topological insulators. *Reviews of Modern Physics* **82**, 3045 (2010).

[111]    D. Wang, L. Kong, P. Fan, H. Chen, S. Zhu, W. Liu, L. Cao, Y. Sun, S. Du, J. Schneeloch, R. Zhong, G. Gu, L. Fu, H. Ding, and H.-J. Gao, Evidence for Majorana bound states in an iron-based superconductor. *Science* **362**, 333 (2018).

[112]    K. Ishida, H. Mukuda, Y. Kitaoka, K. Asayama, Z. Q. Mao, Y. Mori, and Y. Maeno, Spin-triplet superconductivity in Sr2RuO4 identified by 17O Knight shift. *Nature* **396**, 658 (1998).

[113]    K. D. Nelson, Z. Q. Mao, Y. Maeno, and Y. Liu, Odd-Parity Superconductivity in $Sr_2RuO_4$. *Science* **306**, 1151 (2004).

[114]    M. Rice, Superfluid Helium-3 Has a Metallic Partner. *Science* **306**, 1142 (2004).

[115]    I. Žutić and I. Mazin, Phase-Sensitive Tests of the Pairing State Symmetry in $Sr_2RuO_4$. *Physical Review Letters* **95**, 217004 (2005).

[116]    A. N. Petsch, M. Zhu, M. Enderle, Z. Q. Mao, Y. Maeno, I. I. Mazin, and S. M. Hayden, Reduction of the Spin Susceptibility in the Superconducting State of $Sr_2RuO_4$ Observed by Polarized Neutron Scattering. *Physical Review Letters* **125**, 217004 (2020).

[117]    R. Sharma, S. D. Edkins, Z. Wang, A. Kostin, C. Sow, Y. Maeno, A. P. Mackenzie, J. C. S. Davis, and V. Madhavan, Momentum-resolved superconducting energy gaps of $Sr_2RuO_4$ from quasiparticle interference imaging. *Proceedings of the National Academy of Sciences* **117**, 5222 (2020).

[118]    L. Fu and C. L. Kane, Superconducting Proximity Effect and Majorana Fermions at the Surface of a Topological Insulator. *Physical Review Letters* **100**, 096407 (2008).

[119]    Ö. Gül, H. Zhang, J. D. S. Bommer, M. W. A. de Moor, D. Car, S. R. Plissard, E. P. A. M. Bakkers, A. Geresdi, K. Watanabe, T. Taniguchi, and L. P. Kouwenhoven, Ballistic Majorana nanowire devices. *Nature Nanotechnology* **13**, 192 (2018).

[120]    R. M. Lutchyn, E. P. A. M. Bakkers, L. P. Kouwenhoven, P. Krogstrup, C. M. Marcus, and Y. Oreg, Majorana zero modes in superconductor–semiconductor heterostructures. *Nature Reviews Materials* **3**, 52 (2018).

[121]    V. Mourik, K. Zuo, S. M. Frolov, S. R. Plissard, E. P. A. M. Bakkers, and L. P. Kouwenhoven, Signatures of Majorana Fermions in Hybrid Superconductor-Semiconductor Nanowire Devices. *Science* **336**, 1003 (2012).

[122]    M. T. Deng, S. Vaitiekėnas, E. B. Hansen, J. Danon, M. Leijnse, K. Flensberg, J. Nygård, P. Krogstrup, and C. M. Marcus, Majorana bound state in a coupled quantum-dot hybrid-nanowire system. *Science* **354**, 1557 (2016).

[123]    J.-P. Xu, M.-X. Wang, Z. L. Liu, J.-F. Ge, X. Yang, C. Liu, Z. A. Xu, D. Guan, C. L. Gao, D. Qian, Y. Liu, Q.-H. Wang, F.-C. Zhang, Q.-K. Xue, and J.-F. Jia, Experimental Detection of a Majorana Mode in the core of a Magnetic Vortex inside a Topological Insulator-Superconductor $Bi_2Te_3$/$NbSe_2$ Heterostructure. *Physical Review Letters* **114**, 017001 (2015).

[124]    B. E. Feldman, M. T. Randeria, J. Li, S. Jeon, Y. Xie, Z. Wang, I. K. Drozdov, B. Andrei Bernevig, and A. Yazdani, High-resolution studies of the Majorana atomic chain platform. *Nature Physics* **13**, 286 (2017).

[125]  H. Kim, A. Palacio-Morales, T. Posske, L. Rózsa, K. Palotás, L. Szunyogh, M. Thorwart, and R. Wiesendanger, Toward tailoring Majorana bound states in artificially constructed magnetic atom chains on elemental superconductors. *Science Advances* **4**, eaar5251.

[126]  M. Ruby, B. W. Heinrich, Y. Peng, F. von Oppen, and K. J. Franke, Exploring a Proximity-Coupled Co Chain on Pb(110) as a Possible Majorana Platform. *Nano Letters* **17**, 4473 (2017).

[127]  U. Güngördü and A. A. Kovalev, (arXiv:2204.11818, 2022).

[128]  J. Li, H. Chen, I. K. Drozdov, A. Yazdani, B. A. Bernevig, and A. H. MacDonald, Topological superconductivity induced by ferromagnetic metal chains. *Physical Review B* **90**, 235433 (2014).

[129]  S. LaShell, B. A. McDougall, and E. Jensen, Spin Splitting of an Au(111) Surface State Band Observed with Angle Resolved Photoelectron Spectroscopy. *Physical Review Letters* **77**, 3419 (1996).

[130]  G. Nicolay, F. Reinert, S. Hüfner, and P. Blaha, Spin-orbit splitting of the L-gap surface state on Au(111) and Ag(111). *Physical Review B* **65**, 033407 (2001).

[131]  A. C. Potter and P. A. Lee, Topological superconductivity and Majorana fermions in metallic surface states. *Physical Review B* **85**, 094516 (2012).

[132]  J. Alicea, Y. Oreg, G. Refael, F. von Oppen, and M. P. A. Fisher, Non-Abelian statistics and topological quantum information processing in 1D wire networks. *Nature Physics* **7**, 412 (2011).

[133]  P. Bonderson, M. Freedman, and C. Nayak, Measurement-only topological quantum computation via anyonic interferometry. *Annals of Physics* **324**, 787 (2009).

[134]  C. Beenakker, Search for non-Abelian Majorana braiding statistics in superconductors. *SciPost Physics Lecture Notes*, 15 (2020).

[135]  K. Laubscher and J. Klinovaja, Majorana bound states in semiconducting nanostructures. *Journal of Applied Physics* **130**, 081101 (2021).

[136]  T. Zhou, M. C. Dartiailh, K. Sardashti, J. E. Han, A. Matos-Abiague, J. Shabani, and I. Žutić, Fusion of Majorana bound states with mini-gate control in two-dimensional systems. *Nature Communications* **13**, 1738 (2022).

[137]  J. Li, T. Neupert, B. A. Bernevig, and A. Yazdani, Manipulating Majorana zero modes on atomic rings with an external magnetic field. *Nature Communications* **7**, 10395 (2016).

[138]  S. K. Kim, S. Tewari, and Y. Tserkovnyak, Control and braiding of Majorana fermions bound to magnetic domain walls. *Physical Review B* **92**, 020412 (2015).

[139]  U. Güngördü, S. Sandhoefner, and A. A. Kovalev, Stabilization and control of Majorana bound states with elongated skyrmions. *Physical Review B* **97**, 115136 (2018).

[140]  G. L. Fatin, A. Matos-Abiague, B. Scharf, and I. Žutić, Wireless Majorana Bound States: From Magnetic Tunability to Braiding. *Physical Review Letters* **117**, 077002 (2016).

[141]  T. Zhou, N. Mohanta, J. E. Han, A. Matos-Abiague, and I. Žutić, Tunable magnetic textures in spin valves: From spintronics to Majorana bound states. *Physical Review B* **99**, 134505 (2019).

[142]  M. B. Jungfleisch, W. Zhang, R. Winkler, and A. Hoffmann, in *Spin Physics in Semiconductors*, edited by M. I. Dyakonov (Springer International Publishing, Cham, 2017), 355.

[143]  A. D. Kent and D. C. Worledge, A new spin on magnetic memories. *Nature Nanotechnology* **10**, 187 (2015).

[144] E. Y. Tsymbal and I. Žutić, *Spintronics Handbook: Spin Transport and Magnetism, Second Edition* (CRC Press, 2019), Spintronics Handbook: Spin Transport and Magnetism, Second Edition.

[145] N. Mohanta, T. Zhou, J.-W. Xu, J. E. Han, A. D. Kent, J. Shabani, I. Žutić, and A. Matos-Abiague, Electrical Control of Majorana Bound States Using Magnetic Stripes. *Physical Review Applied* **12**, 034048 (2019).

[146] M. M. Desjardins, L. C. Contamin, M. R. Delbecq, M. C. Dartiailh, L. E. Bruhat, T. Cubaynes, J. J. Viennot, F. Mallet, S. Rohart, A. Thiaville, A. Cottet, and T. Kontos, Synthetic spin–orbit interaction for Majorana devices. *Nature Materials* **18**, 1060 (2019).

[147] S.-P. Chiu, C. C. Tsuei, S.-S. Yeh, F.-C. Zhang, S. Kirchner, and J.-J. Lin, Observation of triplet superconductivity in $CoSi_2/TiSi_2$ heterostructures. *Science Advances* **7**, eabg6569 (2021).

[148] Y. Deng, Y. Yu, M. Z. Shi, Z. Guo, Z. Xu, J. Wang, X. H. Chen, and Y. Zhang, Quantum anomalous Hall effect in intrinsic magnetic topological insulator $MnBi_2Te_4$. *Science* **367**, 895 (2020).

[149] J.-X. Yin, W. Ma, T. A. Cochran, X. Xu, S. S. Zhang, H.-J. Tien, N. Shumiya, G. Cheng, K. Jiang, B. Lian, Z. Song, G. Chang, I. Belopolski, D. Multer, M. Litskevich, Z.-J. Cheng, X. P. Yang, B. Swidler, H. Zhou, H. Lin, T. Neupert, Z. Wang, N. Yao, T.-R. Chang, S. Jia, and M. Zahid Hasan, Quantum-limit Chern topological magnetism in $TbMn_6Sn_6$. *Nature* **583**, 533 (2020).

[150] C. Z. Chang, J. Zhang, X. Feng, J. Shen, Z. Zhang, M. Guo, K. Li, Y. Ou, P. Wei, L. L. Wang, Z. Q. Ji, Y. Feng, S. Ji, X. Chen, J. Jia, X. Dai, Z. Fang, S. C. Zhang, K. He, Y. Wang, L. Lu, X. C. Ma, and Q. K. Xue, Experimental observation of the quantum anomalous Hall effect in a magnetic topological insulator. *Science* **340**, 167 (2013).

[151] F. Ando, Y. Miyasaka, T. Li, J. Ishizuka, T. Arakawa, Y. Shiota, T. Moriyama, Y. Yanase, and T. Ono, Observation of superconducting diode effect. *Nature* **584**, 373 (2020).

[152] C. Baumgartner, L. Fuchs, A. Costa, S. Reinhardt, S. Gronin, G. C. Gardner, T. Lindemann, M. J. Manfra, P. E. Faria Junior, D. Kochan, J. Fabian, N. Paradiso, and C. Strunk, Supercurrent rectification and magnetochiral effects in symmetric Josephson junctions. *Nature Nanotechnology* **17**, 39 (2022).

[153] W. Mayer, M. C. Dartiailh, J. Yuan, K. S. Wickramasinghe, E. Rossi, and J. Shabani, Gate controlled anomalous phase shift in Al/InAs Josephson junctions. *Nature Communications* **11**, 212 (2020).

[154] M. C. Dartiailh, W. Mayer, J. Yuan, K. S. Wickramasinghe, A. Matos-Abiague, I. Žutić, and J. Shabani, Phase Signature of Topological Transition in Josephson Junctions. *Physical Review Letters* **126**, 036802 (2021).

[155] A. A. D. Monroe, and I. Žutić, Tunable planar Josephson junctions driven by time-dependent spin-orbit coupling. *Phys. Rev. Applied* **in press** (2022).

[156] O. V. Dobrovolskiy and A. V. Chumak, Nonreciprocal magnon fluxonics upon ferromagnet/superconductor hybrids. *Journal of Magnetism and Magnetic Materials* **543**, 168633 (2022).

**Author biographies and photographs**

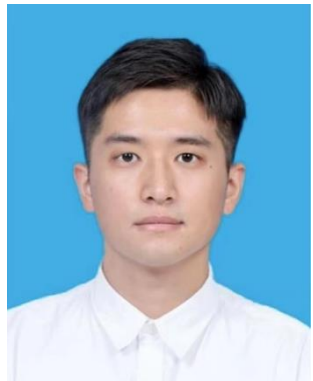

**Dr. Ranran Cai** received his PhD degree in physics at Peking University in 2022. Currently, he is a postdoctoral researcher at USTC. His research interests focused on superconducting spintronics and semiconductor quantum computation.

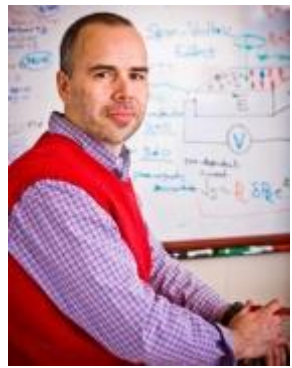

**Prof. Igor Žutić** has received a PhD in physics at the University of Minnesota in 1998. He is a Professor of Physics at the University at Buffalo, the State University of New York and a fellow of the American Physical Society. His work spans topics from unconventional superconductors, Majorana bound states, proximity effects, and two-dimensional materials, to prediction of various spin-based devices, including spin photodiodes, spin solar cells, spin transistors, and spin lasers.

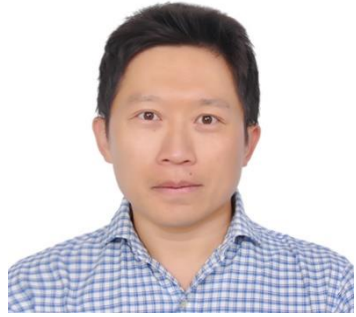

**Prof. Wei Han** received his PhD from the University of California, Riverside, in 2012. He worked at the IBM Almaden Research Center from 2012 to 2014. In 2014, he joined the International Center for Quantum Materials at Peking University. He is currently a tenured associate professor, and his research lab focuses on spintronics in quantum materials.